\documentclass[twocolumn]{aastex701}

\usepackage{graphicx}
\usepackage{booktabs}
\usepackage{multirow}
\usepackage{placeins}
\usepackage{longtable}
\usepackage{array}
\usepackage{tabularx}
\usepackage{float}
\usepackage{xcolor}
\usepackage{subcaption}
\graphicspath{{Figures/}{./}}
\begin{document}

\title{AstroBind: Machine learning prediction of binding energy distributions on interstellar water ice from geometric surface descriptors}

\author[orcid=0009-0005-6138-492X,gname=Aneesa,sname=Ahmad]{Aneesa Ahmad}
\affiliation{School of Physics and Astronomy, University of Leeds, Leeds, UK}
\email[show]{a.ahmad@leeds.ac.uk}

\author[orcid=0000-0001-6078-786X,gname=Catherine,sname=Walsh]{Catherine Walsh}
\affiliation{School of Physics and Astronomy, University of Leeds, Leeds, UK}
\email{c.walsh@leeds.ac.uk}
%%%%%%%%%%%%%%%%%%%%%%%%%%%%%%%%%%%%%%%%%%%%%%%%%%%%%%%%%%%%%%%%%%%%%%%%%%%%%
%%%%%%%%%%%%%%%%%%%%%%%%%%%%%%%%%%%%%%%%%%%%%%%%%%%%%%%%%%%%%%%%%%%%%%%%%%%%%
\begin{abstract}
Binding energies on amorphous solid water regulate molecular desorption in molecular clouds and protoplanetary disks, but computing them from first principles across the heterogeneous binding sites of disordered ice surfaces is computationally prohibitive. We aim to develop a rapid, interpretable method for predicting binding energies on amorphous solid water that captures surface heterogeneity without requiring explicit electronic-structure calculations. We trained a machine-learning model using 27 geometric descriptors of the local ice--adsorbate environment. The model was trained on binding energies for 13 adsorbates on amorphous solid water clusters and evaluated through pooled leave-one-cluster-out validation across 15 ice surfaces. Its transferability was assessed on radical species excluded entirely from training. The model achieves a pooled leave-one-cluster-out coefficient of determination $R^2 = 0.90$ (the fraction of variance in the binding energies captured by the model, where 1 is a perfect fit), and a mean absolute error of 378~K across 15 ice surfaces. It also transfers to radical species withheld from training, with a pooled $R^2 = 0.79$, although its accuracy decreases when the unpaired electron contributes directly to the surface interaction. Our results demonstrate a route towards rapid, distribution-aware parameterisation of binding energies and desorption/diffusion rates in gas--grain astrochemical models, while preserving physical interpretability on heterogeneous ice surfaces.
\end{abstract}
%%%%%%%%%%%%%%%%%%%%%%%%%%%%%%%%%%%%%%%%%%%%%%%%%%%%%%%%%%%%%%%%%%%%%%%%%%%%%
%%%%%%%%%%%%%%%%%%%%%%%%%%%%%%%%%%%%%%%%%%%%%%%%%%%%%%%%%%%%%%%%%%%%%%%%%%%%%
%%%%%%%%%%%%%%%%%%%%%%%%%%%%%%%%%%%%%%%%%%%%%%%%%%%%%%%%%%%%%%%%%%%%%%%%%%%%%
%%%%%%%%%%%%%%%%%%%%%%%%%%%%%%%%%%%%%%%%%%%%%%%%%%%%%%%%%%%%%%%%%%%%%%%%%%%%%
\section{Introduction}
\label{intro}

The chemical complexity inherited by young planets is ultimately governed by processes unfolding at the Angström scale, where molecules adsorb onto and desorb from the disordered surfaces of interstellar ice mantles \citep{_berg_2021, vanDishoeck2014, Cuppen2017}. Binding energies (BEs) on interstellar grain surfaces dictate when and where molecules return to the gas phase, setting thermal desorption fronts, snow-line locations and the balance between gas--phase and grain--surface chemistry \citep{Penteado_2017, minissale_2022}. BEs also control the diffusion rates of species across and within ice mantles thus mediating the reaction rates of grain-surface reactions \citep{ligterink2025}.

Water ice is the dominant mantle component \citep{Boogert_2015, McClure_2023}, making BEs on amorphous solid water (ASW) the most consequential. Experimental temperature-programmed desorption (TPD) measurements \citep{Collings_2004,He_2016} and quantum chemical calculations on model ice surfaces \citep{Das_2018, Ferrero_2020,Sil_2024,Shimonishi_2018} have yielded BE estimates for several tens of species, and community databases now collate recommended values for use in astrochemical models \citep{McElroy2013, minissale_2022, wakelam2015}. However, ASW is intrinsically disordered, presenting a heterogeneous landscape of binding sites that gives rise to broad BE distributions rather than the single values traditionally adopted in gas–grain models \citep{Bovolenta_2022,Grassi_2020,furuya2024, ahmad2026}. Computing these distributions from first principles requires sampling tens to hundreds of adsorbate–surface configurations per species with density functional theory (DFT), each calculation taking hours, making systematic coverage of the molecular inventory prohibitively expensive \citep{Bovolenta_2022,Ferrero_2020}.

Recent work has begun to address this bottleneck with machine learning. \citet{Villadsen_2022} trained models on experimental TPD data to predict single BEs for 114 molecules across surface classes, achieving broad chemical coverage but returning one value per molecule--surface pair rather than a site-resolved distribution. At the other extreme, \citet{Bovolenta_2025} constructed a Gaussian-moment neural network potential for CO on periodic ASW slabs that resolves individual binding sites, but requires retraining for each new adsorbate. \citet{Molpeceres_2020} similarly employed neural network potentials to study nitrogen atom dynamics on ASW. In each case, transferability across chemical species remains limited, and the models rely on electronic--structure information that is itself expensive to generate. Whether local ice-surface geometry alone, without electronic-structure information, contains sufficient information to predict molecular binding has not been systematically tested across chemically diverse adsorbates.

Here we test whether local binding-site geometry alone predicts BEs, without electronic-structure input. Using 27 geometric descriptors of the surface environment, we train a machine-learning (ML) model on BEs for 13 adsorbates across ASW clusters and assess its transfer to radical species withheld from training. We propagate the predicted BE distributions through the Polanyi–Wigner equation to recover TPD peak desorption temperatures. Our aim is a fast, interpretable, distribution-aware route to parameterising BEs in gas–grain astrochemical models.
%%%%%%%%%%%%%%%%%%%%%%%%%%%%%%%%%%%%%%%%%%%%%%%%%%%%%%%%%%%%%%%%%%%%%%%%%%%%%
%%%%%%%%%%%%%%%%%%%%%%%%%%%%%%%%%%%%%%%%%%%%%%%%%%%%%%%%%%%%%%%%%%%%%%%%%%%%%
\section{Methods}
\subsection{Dataset}
\label{dataset}

Binding energy (BE) data were taken from \citet{ahmad2026}, who computed adsorption energies using DFT for 13 molecular species on amorphous solid water (ASW) clusters composed of 12 water molecules (ASW$_{12}$). Eight closed-shell (CH$_3$OH, H$_2$O, CO, CO$_2$, H$_2$CO, HCOOH, CH$_4$, C$_2$H$_2$) and five radical species (OH, HCO, CH$_3$O, CH$_2$OH, CH$_3$) were studied. BEs were calculated using eight base exchange--correlation functionals spanning three classes: hybrid meta-GGA (M06-HF, \citealt{zhao2006}; MPWB1K, \citealt{Zhao2004}; PW6B95, \citealt{zhao2005}; TPSSh, \citealt{tao2003, staroverov2003}); range-separated hybrid ($\omega$B97M-V, \citealt{Mardirossian2016}; $\omega$B97X-V, \citealt{Mardirossian2014}; $\omega$PBE, \citealt{vydrov2006}); and GGA (B97-2, \citealt{Wilson2001}). Grimme's D3BJ dispersion correction \citep{Grimme2010} was applied to the five functionals lacking built-in dispersion (B97-2, MPWB1K, PW6B95, TPSSh, and $\omega$PBE), yielding 13 functional variants in total; $\omega$B97M-V and $\omega$B97X-V already incorporate dispersion, and M06-HF was used without an added correction.

Up to 11 functionals were available per structure (median 4), and the ensemble mean across all available functionals was used as the target variable to mitigate individual functional bias. The final dataset contained 1,429 adsorbate--surface configurations distributed across 15 distinct ASW$_{12}$ cluster geometries, with between 99 (CO and CH$_2$OH) and 153 (CH$_3$OH) structures per adsorbate. BEs extend up to approximately 9,600~K, reflecting the diversity of binding sites on amorphous ice. 
%All structures and binding-energy distributions are publicly available on Zenodo (\textcolor{red}{DOI to be added}).
%%%%%%%%%%%%%%%%%%%%%%%%%%%%%%%
\subsection{Feature Extraction}
\label{feature_extrac}
For each adsorbate--surface configuration, we extracted 27 geometric descriptors characterising the local binding environment. These comprise hydrogen-bonding (H bonding) metrics (numbers of donated, accepted, and total hydrogen bonds (H-bonds); mean and minimum H-bond distance; mean H-bond angle); dangling-OH proximity, given by the number of dangling OH groups lying within 5\,\AA{} of the adsorbate together with the nearest and mean adsorbate--dangling-OH distances; adsorbate--surface distances (minimum, mean, and centre-of-mass-to-surface); local water density within 3, 5, and 7\,\AA{} shells; coordination number; surface contact area; adsorbate and surface centre-of-mass coordinates; surface H-bond network density; and atom counts.

H-bonding descriptors are undefined for configurations in which no hydrogen bond is formed. For these cases, the corresponding geometric features were set to zero and binary indicator variables were used to encode the physical absence of an H-bond. In particular, the indicator associated with the average H-bond distance is equivalent to a binary descriptor specifying whether a configuration forms no hydrogen bond. This yields six additional binary features.

Adsorbate identity was encoded using a one-hot scheme, in which each of the 13 species is represented by a 13-element binary vector with a single element set to one (identifying that species) and the remainder set to zero. This treats adsorbate identity as a categorical variable, imposing no artificial ordering or numerical relationship between species. Combined with the 33 scalar geometric features (with vector-valued quantities such as centre-of-mass coordinates expanded into their components), this gives a total of 46 input features per structure (Appendix Table~\ref{tab:feature_definitions}, Figure~\ref{fig:feature_importance_all}). The ratio of training samples to input features ($1{,}429{:}46 \approx 31{:}1$) comfortably exceeds the minimum thresholds recommended for descriptor-based regression models, reducing the risk of chance correlations arising from the dimensionality of the feature space.
%%%%%%%%%%%%%%%%%%%%%%%%%%%%%
\subsection{Model Training}
\label{model_training}
We modelled BEs with gradient-boosted decision trees, an ensemble method that builds a sequence of shallow regression trees in which each new tree is fitted to the residual errors of those before it; the final prediction is the sum of all tree outputs, so the model corrects its own errors incrementally. We used the XGBoost implementation \citep{chen2016} (v2.1.4) \footnote{\url{https://github.com/dmlc/xgboost}}, which is fast and has built-in protection against overfitting. The model's hyperparameters (the configuration values we set before training, rather than quantities the model learns from the data) were: 100 estimators, learning rate 0.1, maximum depth 5, minimum child weight 3, subsample fraction 0.8, and column subsample fraction 0.8. The three most influential are the number of estimators (how many trees the ensemble builds), the maximum depth (how many successive splits each tree may make, which sets how complex a pattern it can fit), and the learning rate (how strongly each new tree adjusts the running prediction). Varying each of these around the chosen values changes performance only slightly (Appendix Table~\ref{tab:hyperparam_sensitivity}), confirming that the configuration is robust rather than finely tuned. All continuous features were standardised to zero mean and unit variance using the StandardScaler routine from scikit-learn \citep{pedregosa2011},\footnote{\url{https://scikit-learn.org/stable/modules/generated/sklearn.preprocessing.StandardScaler.html}} fitted on the training set only to prevent data leakage.

To confirm that the chosen architecture was appropriate, we compared four regression methods: linear regression, ridge regression \citep{hoerl1970}, random forest \citep{breiman2001}, and XGBoost (Appendix Table~\ref{tab:model_comparison}). Linear and ridge regression serve as baselines representing purely linear mappings from geometric descriptors to the BE; both achieve competitive overall accuracy (the coefficient of determination $R^2 = 0.88$, i.e.\ the model reproduces 88 per cent of the variance in the BE, where $R^2 = 1$ denotes a perfect fit) but cannot capture non-linear dependencies or conditional interactions between features. For example, the influence of H-bond geometry on the BE is inherently adsorbate-dependent, vanishing entirely for non-H bonding species. Random forest models capture such non-linearities through independent decision trees but lack the sequential error-correction of boosted methods. XGBoost achieved the highest test-set accuracy ($R^2 = 0.89$) and is robust to multicollinearity (strong correlations between input features, Appendix Figure~\ref{fig:correl_matrix}), which is present in the geometric descriptor set (e.g., the mean dangling-OH distance and the mean adsorbate--surface distance are correlated at $r = 0.89$). Hyperparameter tuning via randomised-search cross-validation, in which 80 randomly sampled hyperparameter combinations are each scored by 5-fold cross-validation (the training data split into five parts, with each held out for validation in turn), yielded only marginal improvement ($R^2 = 0.88$–$0.89$), suggesting that the baseline configuration was already close to optimal.

We also benchmarked Gaussian process regression (GPR) and kernel ridge regression, as GPR was employed by \citet{Villadsen_2022} for a related study. On the same 80/20 cluster-based split, GPR achieves $R^2 = 0.87$ and a mean absolute error (MAE) of $454$~K, compared with $R^2 = 0.90$ and MAE $= 380$~K for XGBoost (Appendix Table~\ref{tab:model_comparison}); the MAE is the average absolute difference between predicted and reference BEs, in kelvin, so lower values indicate more accurate predictions. The lower performance of the kernel methods likely reflects their assumption of a single global length scale, which is poorly suited to the heterogeneous binding regimes present in the data. GPR additionally scales as ${O}(n^3)$ with training-set size $n$; that is, its computational cost grows with the cube of the number of training samples, which limits its applicability to larger datasets.
%%%%%%%%%%%%%%%%%%%%%%%%%%%%%%%%%%%%%%
\subsection{Cross-validation strategy}
\label{CV_strat}

Data were split at the cluster level to ensure the model generalises to unseen ice-surface topologies. In the primary evaluation, 12 ASW$_{12}$ clusters were used for training and 3 (ASW$_{12}$\_01, ASW$_{12}$\_10, ASW$_{12}$\_14) were held out for testing, giving an 80/20 split (1,162 training, 267 test structures). To assess robustness, leave-one-cluster-out (LOCO) cross-validation was performed across all 15 ASW$_{12}$ clusters. In each of the 15 folds, one cluster was held out as the test set and the model was retrained on the remaining 14, so that all 1,429 ASW$_{12}$ structures received exactly one blind prediction. The pooled LOCO $R^2$ was 0.90 (MAE $= 378$~K), with the per-cluster breakdown given in Appendix Table~\ref{tab:loco_clusters}; this confirms that the single-split result is representative and not an artefact of the particular cluster partition.

To assess dataset sufficiency, we constructed a learning curve by training on progressively larger random subsets of clusters (5--15) and evaluating pooled LOCO performance at each step (Appendix Table~\ref{tab:learning_curve}). Test $R^2$ reaches 0.88 with just five clusters ($\sim$490 structures) and plateaus at 0.89--0.90 from seven clusters onward, while the inter-subset standard deviation decreases from $\pm 0.04$ to $\pm 0.001$. This indicates that the current dataset captures the geometric diversity relevant to binding, and that additional clusters would yield diminishing returns.

As a control for chance correlations, we repeated the full training procedure with randomly shuffled BE labels (10 independent shuffles). All runs yielded negative $R^2$ values (range $-0.10$ to $-0.17$; mean $-0.13$), supporting the conclusion that the model captures genuine structure-property relationships rather than chance correlations.
%%%%%%%%%%%%%%%%%%%%%%%%%%%%%%%%%%%%%%%%%%%
\subsection{Radical withholding experiment}
\label{rad_witholding_expt}

To test whether geometric features learned from closed-shell species transfer to chemically distinct adsorbates, a separate model was trained using only the eight non-radical species (748 structures across the 12 training clusters). This model was evaluated on the five radical species (CH$_3$, HCO, OH, CH$_3$O, CH$_2$OH) using the same held-out test clusters as the primary evaluation. No radical data were seen during training. The pooled $R^2$ of 0.79 across all radical predictions reflects primarily the model's ability to rank radical species by binding strength; within-species accuracy varies considerably depending on whether the unpaired electron participates directly in the surface interaction (Table~\ref{tab:summary_metrics}, Figure~\ref{fig:LOCO_rad}b).
%%%%%%%%%%%%%%%%%%%%%%%%%%%%%%%%%%%%%%%%%%%%%%%%%%%%%%%%%%%%%%%%%%%%%%%%%%%%%
%%%%%%%%%%%%%%%%%%%%%%%%%%%%%%%%%%%%%%%%%%%%%%%%%%%%%%%%%%%%%%%%%%%%%%%%%%%%%
\section{Results}
\label{results}

\begin{figure*}
  \centering
  \includegraphics[width=\textwidth]{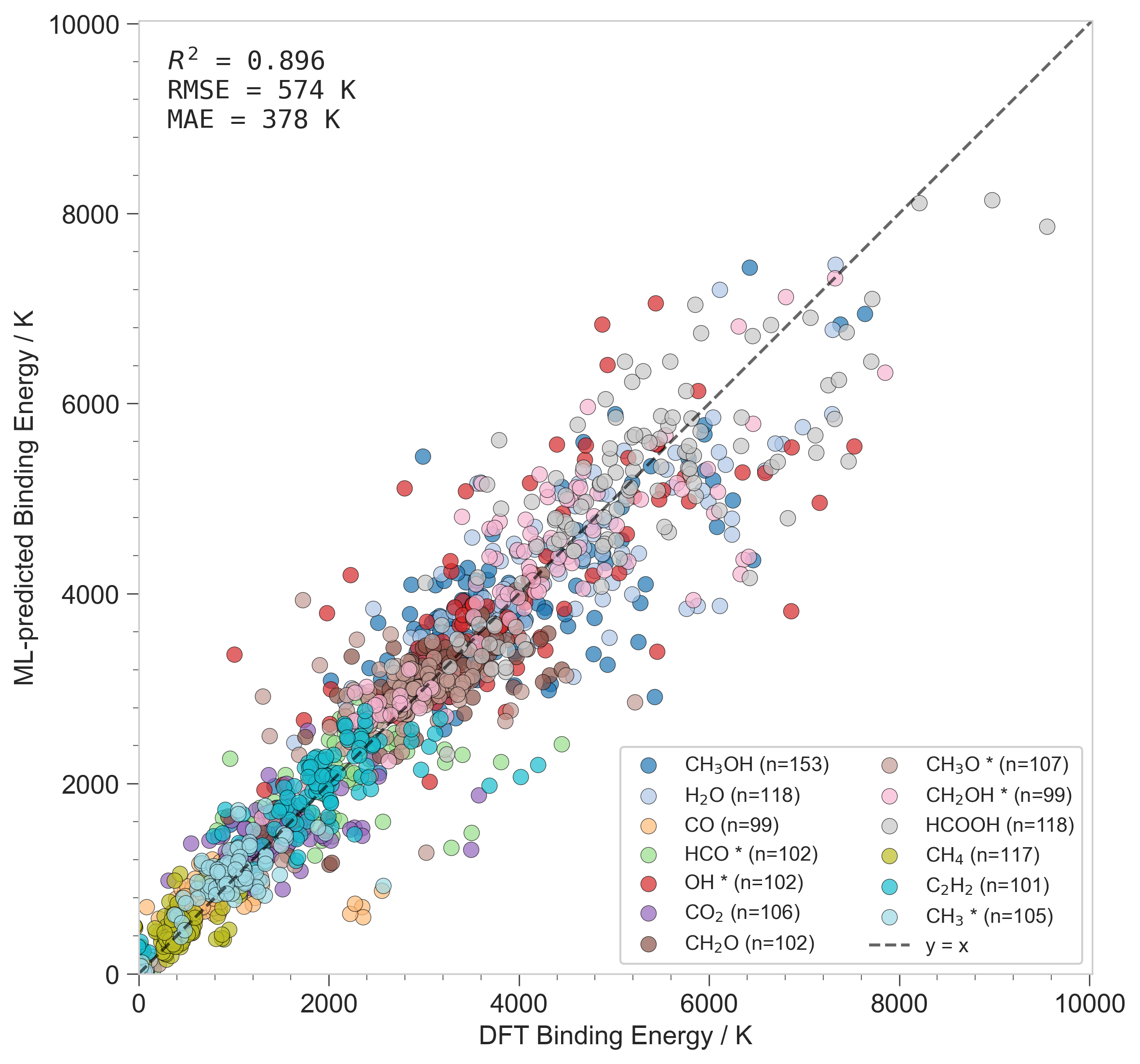}
  \caption{LOCO-CV predicted vs DFT binding energies for 13 adsorbate species on ASW$_{12}$ clusters. Each point represents a single adsorbate–surface configuration, coloured by species identity; asterisks denote radical species. Pooled metrics across all leave-one-cluster-out folds: R$^2$ = 0.896, RMSE = 574 K, MAE = 378 K.}
  \label{fig:BEs_scatter}
\end{figure*}

We trained a gradient-boosted decision tree model (XGBoost; \citealt{chen2016}) to predict the BEs of 13 adsorbate species on ASW$_{12}$ clusters using 27 geometric descriptors of the binding site environment. The model was evaluated using a cluster-based train/test split (12/3 clusters) and validated with leave-one-cluster-out (LOCO) cross-validation across all 15 clusters. A summary of the per-adsorbate binding energy statistics and model performance metrics is given in Table~\ref{tab:summary_metrics} and Figure~\ref{fig:BEs_scatter}.

\begin{table*}[t]
\centering
\caption{Summary statistics and model performance. DFT $\mu$ and $\sigma$ are the mean and standard deviation from Gaussian fits to the binding energy distributions \citep{ahmad2026}. ML $\mu$ and $\sigma$ are the arithmetic mean and standard deviation of the ML-prediction on the test set (3 ASW$_{12}$ clusters). LOCO R$^2$ and MAE are pooled metrics from 15-fold leave-one-cluster-out cross-validation across all ASW$_{12}$ binding sites. For radical species (HCO, OH, CH$_3$O, CH$_2$OH, CH$_3$), the withheld MAE reports the error from a model trained exclusively on non-radical species.}
\label{tab:summary_metrics}
\renewcommand{\arraystretch}{1.25}
\begin{tabular*}{\textwidth}{@{\extracolsep{\fill}}l cc cc cc c@{}}
\toprule
 & \multicolumn{2}{c}{DFT (K)}
 & \multicolumn{2}{c}{ML (K)}
 & \multicolumn{2}{c}{LOCO}
 & Withheld \\
\cmidrule(lr){2-3}
\cmidrule(lr){4-5}
\cmidrule(lr){6-7}
\cmidrule(l){8-8}
Species & $\mu$ & $\sigma$ & $\mu$ & $\sigma$ & $R^{2}$ & MAE (K) & MAE (K) \\
\midrule
CH$_3$OH   & 3344 & 1005 & 3879 & 1033 & 0.67    & 475 & \dots \\
H$_2$O     & 3376 & 1081 & 4011 &  684 & 0.74    & 518 & \dots \\
CO         &  624 &  201 &  655 &  125 & 0.07    & 264 & \dots \\
HCO        & 1699 &  572 & 1894 &  576 & 0.66    & 325 & 547 \\
OH         & 3041 & 1145 & 4015 &  926 & 0.53    & 595 & 764 \\
CO$_2$     & 1546 &  538 & 1505 &  282 & 0.53    & 280 & \dots \\
H$_2$CO    & 2583 &  634 & 2940 &  678 & 0.68    & 310 & \dots \\
CH$_3$O    & 2281 &  610 & 2788 &  398 & 0.20    & 389 & 525 \\
CH$_2$OH   & 3602 & 1293 & 4034 & 1021 & 0.82    & 463 & 528 \\
HCOOH      & 4714 & 1397 & 5105 & 1068 & 0.69    & 578 & \dots \\
CH$_4$     &  418 &  130 &  473 &  168 & $-0.02$ & 152 & \dots \\
C$_2$H$_2$ & 1734 &  545 & 1962 &  345 & 0.62    & 318 & \dots \\
CH$_3$     &  778 &  251 & 1031 &  274 & 0.64    & 192 & 198 \\
\bottomrule
\end{tabular*}
\end{table*}
%%%%%%%%%%%%%%%

\begin{figure*}[t]
  \centering
  \includegraphics[width=0.85\textwidth,height=0.60\textheight,keepaspectratio]{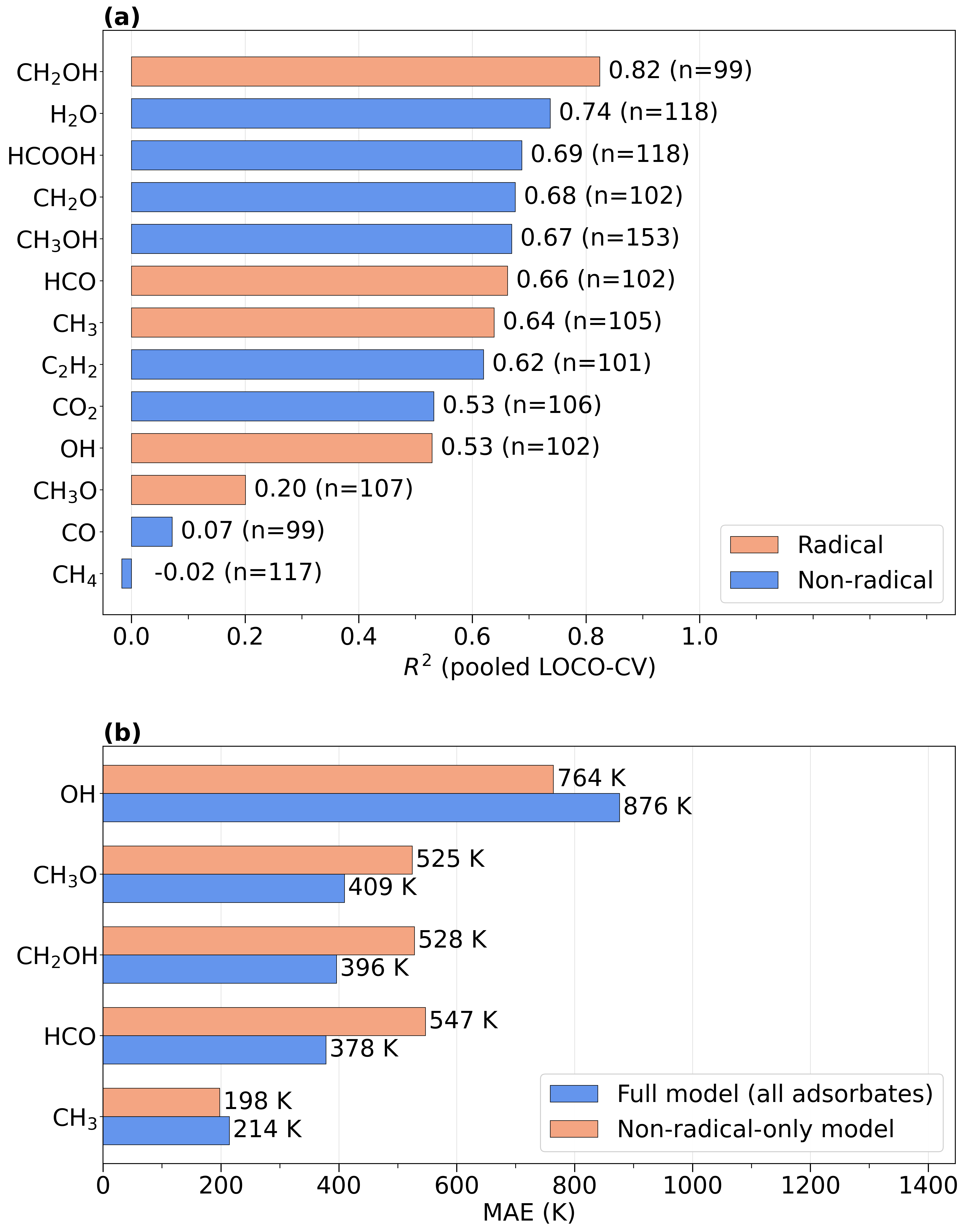}
  \caption{Per-adsorbate model performance and radical transferability. (a) Pooled LOCO-CV $R^2$ for each species. $n$ is the number of binding sites. (b) MAE (K) for radical species from the full model and the non-radical-only model.}
  \label{fig:LOCO_rad}
\end{figure*}

The model captures the species-to-species hierarchy of BEs on amorphous water ice, correctly ranking hydrogen-bonding species (HCOOH, CH$_2$OH, H$_2$O) as the strongest binders and dispersion-bound volatiles (CO, CH$_4$) as the weakest. On the held-out test set (3 ASW$_{12}$ clusters, $n = 267$), the model achieves an overall $R^2 = 0.89$, RMSE $= 593$~K, and MAE $= 387$~K across all thirteen species. The predicted BEs track the $y = x$ line from approximately 200~K to 9{,}600~K, with scatter increasing at higher BEs where the diversity of binding modes is greatest. LOCO cross-validation, in which each of the 15 ASW$_{12}$ clusters is held out in turn, yields a pooled $R^2 = 0.90$ and MAE $= 378$~K (Figure~\ref{fig:BEs_scatter}), confirming that the single-split result is robust and not an artefact of the particular cluster assignment (Figure~\ref{fig:LOCO_rad}a).

A linear fit to the predicted-versus-actual values (Figure~\ref{fig:LOCO_rad}a) gives a calibration slope of 0.89, meaning the model slightly narrows the spread of predicted BEs relative to the DFT reference ($\sigma_\mathrm{pred}/\sigma_\mathrm{actual} = 0.94$, a compression of about 6\%). This mild attenuation is a familiar effect of regression to the mean in finite training sets. The errors are not spread evenly across the BE range: they are largest for the most weakly bound quartile (MAE $= 665$~K), about three times those for the most strongly bound quartile (MAE $= 212$~K). Weakly bound species sit in dispersion-dominated sites, which vary more in geometry and are less well captured by the geometric descriptors, so the model predicts them less accurately.

To attach an uncertainty estimate to each prediction, we trained a bootstrap ensemble of 50 XGBoost models, each fitted to a resample of the training data drawn with replacement at the cluster level, and used the spread (standard deviation) of their predictions as the uncertainty for each binding site. The mean predicted uncertainty is 225~K (range 50--609~K). Larger predicted uncertainties tend to accompany larger actual errors: the two are positively rank-correlated (Spearman $\rho = 0.365$, $p = 1.4\times10^{-12}$). Spearman's $\rho$ measures whether two quantities increase together in rank order, on a scale from $-1$ to $+1$, so this moderate positive value means a higher predicted uncertainty generally goes with a higher error, though not perfectly. Grouping predictions by their ensemble spread, the observed RMSE rises from 268~K in the most confident fifth of predictions to 706~K in the least confident (Appendix Table~\ref{tab:uncertainty_calibration}). The ensemble tends to underestimate predictive uncertainty, as actual errors exceed predicted uncertainties in every bin. Despite this, the ranking is preserved, allowing the model to reliably identify its least trustworthy predictions.
%%%%%%%%%%%%%%%%%%%%%%%%%%%%%%%%%%%%%%%%%%%%%%
\subsection{Per-adsorbate predictive accuracy}
\label{per_ads_accuracy}

\begin{figure*}[p]
  \centering
  \includegraphics[width=\textwidth,height=0.95\textheight,keepaspectratio]{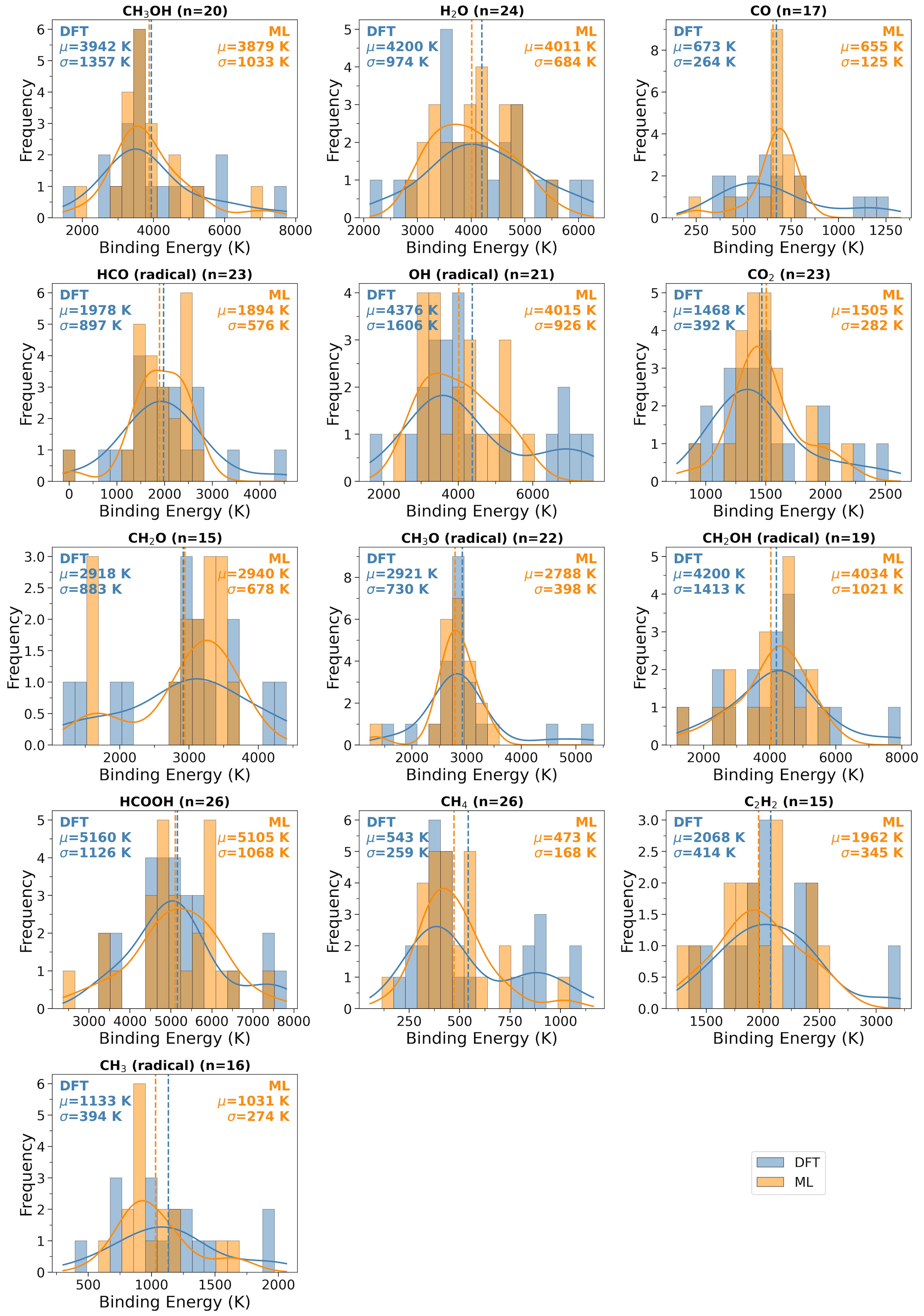}
  \caption{ML-predicted versus DFT binding energy distributions for 13 adsorbates on amorphous water ice. Blue and orange histograms show DFT-calculated and ML-predicted binding energies, respectively, for the held-out test set (3 ASW$_{12}$ clusters). Dashed vertical lines indicate mean values; solid curves are kernel density estimates. Radical species are labelled. $\mu$ and $\sigma$ denote the mean and standard deviation of each distribution, and $n$ is the number of test structures for each species.}
  \label{fig:dftvsML}
\end{figure*}

The per-adsorbate predicted and DFT BE distributions are shown in Figure~\ref{fig:dftvsML}, with pooled LOCO R$^2$ values for each species shown in Figure~\ref{fig:LOCO_rad}a. Predictive accuracy varies systematically with the dominant binding mechanism, revealing three distinct regimes.

Species that bind primarily through hydrogen bonds achieve the highest predictive accuracy: CH$_2$OH (R$^2$ = 0.82), H$_2$O (0.74), and HCOOH (0.69). CH$_3$OH (0.67) and H$_2$CO (0.68) show moderate performance, likely reflecting the greater diversity of binding configurations accessible to these species on the ASW$_{12}$ surface. For all hydrogen-bonding species, the model captures the mean and general shape of the BE distribution (Figure~\ref{fig:dftvsML}), although the predicted distributions are consistently narrower than the DFT reference (Table~\ref{tab:summary_metrics}), a consequence of regression towards the training mean.

Dispersion-dominated species exhibit markedly lower predictive accuracy. CO (R$^2$ = 0.07) and CH$_4$ (R$^2 = -0.02$) bind through weak van der Waals interactions with no directional hydrogen-bonding character. Their BE distributions are narrow ($\sigma$ = 130–201 K), and the geometric descriptors, which primarily encode hydrogen-bond geometry and local water density carry little information about the strength of dispersion interactions at individual binding sites.

Open-shell radical species show intermediate and heterogeneous performance. CH$_2$OH achieves the highest per-species R$^2$ (0.82) despite being a radical, because its binding is mediated by its OH moiety through conventional hydrogen bonding, with the unpaired electron localised on carbon. In contrast, CH$_3$O (R$^2$ = 0.20) and OH (R$^2$ = 0.53), for which the unpaired electron resides on the binding atom, are poorly described by geometric features alone. The limiting factor is therefore not radical character per se, but whether the unpaired electron directly participates in the surface interaction.
%%%%%%%%%%%%%%%%%%%%%%%%%%%%%%%
\subsection{Feature importance}
\label{feature_imp}

Appendix Figure~\ref{fig:feature_importance_all} shows the highest-ranking physical descriptors by XGBoost gain-based feature importance in descending order, including binary H-bond indicators and adsorbate one-hot encodings. The strongest continuous geometric feature is the minimum adsorbate--surface distance (9.7\% of total model gain). H-bonding descriptors also contribute substantially, with the minimum H-bond distance (5.5\%), average H-bond angle (3.4\%), and number of donated H-bonds (1.9\%) among the most informative. Broader measures of the local ice environment, including water density, dangling OH counts, and centre-of-mass position, contribute less, consistent with adsorption being governed primarily by the local binding environment.

The dominant feature is the binary descriptor indicating that no H-bond is formed, which contributes 63.4\% of the total model gain. This descriptor arises from the indicator used when the average H-bond distance is undefined, but has a direct physical interpretation: it separates configurations that do not form H-bonds from those that do. BEs differ strongly between these groups, with mean BEs of 1043~K and 3805~K, respectively. Once this coarse separation is established, the continuous geometric descriptors provide the site-level resolution within each binding regime.

\begin{figure*}[t]
  \centering
  \includegraphics[width=\textwidth,height=0.9\textheight,keepaspectratio]{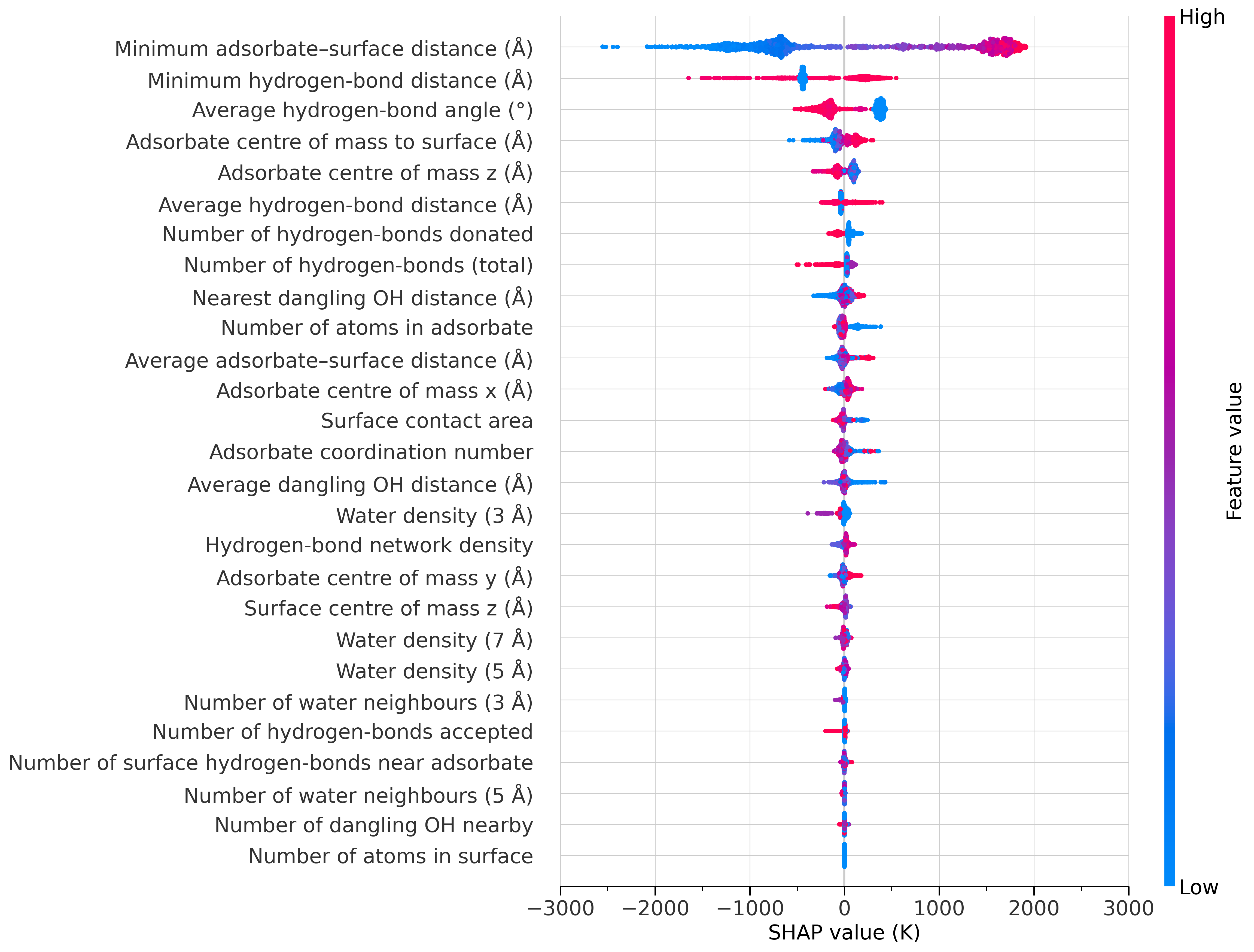}
  \caption{SHAP beeswarm plot for all 27 geometric descriptors used by the XGBoost model. Each point represents one adsorbate--cluster structure ($n = 1{,}429$); the horizontal position gives the SHAP value (contribution to the predicted binding energy, in K), and colour indicates the feature value (blue = low, red = high). Features are ranked by mean $|\mathrm{SHAP}|$. The minimum adsorbate--surface distance dominates (mean $|\mathrm{SHAP}| = 1{,}111$~K), with low values pushing predictions toward stronger binding, as expected. The minimum H-bond distance and average H-bond angle are the next most influential features (383~K and 271~K, respectively), confirming that H-bond geometry is the primary driver of predicted binding energies beyond simple proximity.}
  \label{fig:SHAP}
\end{figure*}

To move beyond the global ranking provided by gain-based importance and examine how individual features influence predictions directionally, we computed SHAP (SHapley Additive exPlanations) values for every prediction in the dataset \citep{lundberg2017}. For each individual prediction, SHAP assigns every feature a value showing how much it pushed that prediction above or below the model's average output: the sign gives the direction (raising or lowering the predicted BE) and the magnitude gives the strength, in the same units as the target (K). These values come from Shapley values in cooperative game theory, a principled way of fairly splitting the credit for an outcome among the contributors, here the input features. The global mean SHAP ranking is broadly consistent with the gain-based ordering: the minimum adsorbate--surface distance dominates (1{,}111~K), followed by the minimum H-bond distance (383~K) and the average H-bond angle (Figure~\ref{fig:SHAP}). However, SHAP additionally reveals the direction and magnitude of each feature's contribution to individual predictions, enabling a regime-resolved analysis that gain-based importance cannot provide.

Stratifying SHAP values by binding strength reveals pronounced regime dependence
in how geometric features drive predictions. For strongly bound sites
($\mathrm{BE} > 7{,}216$~K), the mean $|\mathrm{SHAP}|$ contribution of the
minimum H-bond distance increases by a factor of 3.9 relative to weakly bound
sites ($\mathrm{BE} < 4{,}811$~K), from 340~K to 1{,}312~K, confirming that
H-bond geometry is the primary determinant of site-to-site variation in the
high-BE tail. The total number of H-bonds shows an even sharper contrast,
rising from 52~K to 140~K, indicating that the model has learned that
multidentate H-bonding drives the strongest binding configurations. At
weakly bound, dispersion-dominated sites, these same features contribute
minimally, and the minimum adsorbate--surface distance alone accounts for most
of the predicted variation. This regime-dependent behaviour is consistent with the model capturing physically distinct binding regimes rather than relying on a single global mapping.

Comparing SHAP attributions for radical and non-radical species reveals that the same geometric features dominate both subsets with similar magnitudes, indicating that geometric binding principles transfer across chemical classes. The differentiation between radicals and non-radicals arises primarily through the adsorbate identity encodings rather than through different geometric dependencies, illustrating that a model trained exclusively on non-radical species captures the broad ordering of radical binding strengths.
%%%%%%%%%%%%%%%%%%%%%%%%%%%%%
\subsection{Radical transferability}
\label{rad_transfer}

To test whether geometric binding principles learned from closed-shell species generalise to radicals, we trained a second model exclusively on the eight non-radical adsorbates and evaluated its predictions for the five radical species on the held-out test clusters (Figure~\ref{fig:LOCO_rad}b). The non-radical-only model predicts radical BEs with MAE values of 198–764 K (Table~\ref{tab:summary_metrics}). The degradation relative to the full model varies considerably across species: CH$_3$ and CH$_2$OH show modest increases of 3\% and 14\%, respectively, while CH$_3$O (35\%) and HCO (68\%) degrade substantially. Although absolute errors remain large for OH (764 K) and HCO (547 K), the model captures the broad ordering of radical binding strengths without explicit radical training data. Geometric features therefore encode partial information about radical adsorption, but electronic effects associated with unpaired electrons limit the accuracy achievable using geometric descriptors alone.
%%%%%%%%%%%%%%%%%%%%%%%%%%%%%%%%%%%%%%%%%%%%%%%%%%%%%%%%%%%%%%%%%%%%%%%%%%%%%
%%%%%%%%%%%%%%%%%%%%%%%%%%%%%%%%%%%%%%%%%%%%%%%%%%%%%%%%%%%%%%%%%%%%%%%%%%%%%
\section{Discussion}
\label{disc}

We adopted a descriptor-based gradient-boosted decision-tree model (XGBoost) rather than data-hungry deep learning architectures because, with 1,429 training structures, boosted trees are competitive on tabular data of this scale \citep{grinsztajn2022,shwartzziv2021,deng2023} and the resulting feature attributions provide direct mechanistic insight. The trade-off is intrinsic: the model cannot learn physics absent from the descriptor set, which is reflected in its regime-dependent performance.

This framework complements existing ML strategies for astrochemical BEs. \citet{Bovolenta_2025} trained a Gaussian-moment neural network on 8,321 DFT energies to construct a CO potential energy surface on periodic ASW slabs, enabling binding-site discovery but requiring retraining for each new adsorbate. \citet{Villadsen_2022} used experimental TPD data to predict single BEs for 114 molecules across surface classes ($R^2=0.96$), achieving broad chemical coverage but returning one BE per molecule–surface pair. Our model operates at the binding-site level, predicting how binding varies across adsorption geometries on amorphous water ice and thereby producing full BE distributions. These approaches are complementary. \citet{Bovolenta_2025} provide high-accuracy, single-species energetics, \citet{Villadsen_2022} achieve broad chemical coverage across surface classes, and our model provides site-resolved BE distributions across multiple adsorbates.

Several limitations define the scope of the present model. First, the model is trained on ASW$_{12}$ clusters, which are smaller than the W22 clusters and periodic slabs employed in recent studies \citep{Ferrero_2020, Bovolenta_2022, Bovolenta_2025}. Absolute DFT BEs on small clusters can deviate from laboratory-derived values and from calculations on larger ice models \citep{Penteado_2017, minissale_2022, Ferrero_2020, Bovolenta_2022}, and the model should therefore be interpreted as reproducing the DFT reference landscape rather than providing direct experimental estimates. However, the model does not predict BEs from cluster size; it predicts them from local geometric descriptors specifically H-bond distances and angles, adsorbate--surface proximity, and coordination number that characterise the binding environment within the first coordination shell ($\sim$3--5~\AA{}). The SHAP analysis (Figure~\ref{fig:SHAP}) confirms that these near-field features dominate the learned mapping: the minimum adsorbate--surface distance (mean $|\mathrm{SHAP}| = 1{,}111$~K), minimum H-bond distance (383~K), and average H-bond angle (271~K) collectively account for the majority of the predicted BE variation, while broader measures of the ice environment beyond the nearest few \AA{} contribute minimally. The physical relationships encoded by these descriptors, including shorter H-bonds yielding stronger binding, multidentate coordination stabilising the adsorbate, and closer surface contact increasing interaction strength, are expected to remain relevant across different ice-model sizes. What larger ice models would provide is a wider diversity of local binding environments, including deeply buried pores and extended flat terraces, that may not be fully represented among the 15 ASW$_{12}$ clusters sampled here. The LOCO validation ($R^2 = 0.90$ across 15 structurally distinct clusters) demonstrates that the learned geometric correlations transfer across the range of binding-site morphologies present in the ASW$_{12}$ dataset; testing whether this transferability extends to larger clusters and periodic slabs is a natural next step. 

Second, performance collapses for dispersion-dominated adsorbates (CO: $R^2 = 0.07$; CH$_4$: $R^2 = -0.02$), consistent with the absence of electronic information: weak physisorption depends on polarisability and subtle electron-density overlap that distances and angles do not encode. A related limitation applies to radical species: for radicals whose unpaired electron directly participates in the surface interaction (CH$_3$O, OH), missing spin-dependent exchange and charge-transfer contributions impose an accuracy ceiling that purely geometric representations cannot overcome. 

Third, although the model gives a single predicted value per site rather than a full probability distribution, a 50-model bootstrap ensemble supplies a per-site uncertainty estimate. These estimates rise with the true errors (Spearman $\rho = 0.365$, $p = 1.4\times10^{-12}$), so the model reliably separates its confident predictions from its uncertain ones. The ensemble tends to underestimate predictive uncertainty, with actual errors exceeding predicted uncertainties across the calibration bins. Conformal prediction, which uses a separate calibration set to convert uncertainty estimates into error bars with a known success rate (e.g.\ capturing the true value 90\% of the time), could tighten these intervals in future work.

Finally, adsorbate identity is represented by one-hot encoding, which limits the model to the 13 species in the training set. The radical withholding experiment (Section~\ref{rad_witholding_expt}) demonstrates that the geometric descriptors alone capture the broad ordering and approximate distribution shapes for unseen species, but precise site-level predictions require the species-specific calibration provided by the identity encoding. Replacing the categorical representation with continuous molecular descriptors (e.g.\ dipole moment, polarisability, number of H-bond donors) would retain this calibration while enabling generalisation to new adsorbates, as demonstrated by \citealt{Villadsen_2022}. These limitations reflect the information available to a geometry-only representation and indicate where low-cost electronic descriptors, such as partial charges, frontier orbital energies, or spin densities, could be most valuable.

Once trained, the model predicts without further DFT. For any of the 13 trained species, the user supplies a new ASW$_{12}$ cluster and receives a full per-site BE distribution in one forward pass, with no DFT sampling of individual sites. This points to a practical use case: compute DFT BEs for a representative set of clusters, train the model, then predict per-site distributions across many more clusters than direct DFT could afford. Because each prediction is a single forward pass, the DFT effort that would otherwise scale with the number of clusters and sites collapses to a fixed up-front training cost, after which a more complete, better-converged BE distribution can be assembled at negligible additional expense. DFT remains necessary to generate the reference data used for training. For a trained species, no additional DFT calculations are required for prediction, whereas a new adsorbate currently requires species-specific DFT data because of the one-hot encoding. Extending this transferability so that future versions predict unseen species directly, without species specific DFT, is a natural next step and would follow from replacing the one-hot identity with continuous molecular descriptors.

Despite these limits, the model captures the physics that matters most for astrochemical models. H-bonding species dominate the ice mantle and drive the thermal desorption that sets snow-line locations and gas-phase abundances \citep{_berg_2021, minissale_2022}. For these species, the predicted BE distributions capture the site-to-site variation that single value parameterisations miss. Most gas--grain models (e.g.\ \citealt{holdship2017}; \citealt{ruaud2016}) assign each species one recommended BE from databases such as UMIST \citep{McElroy2013} and KIDA \citep{wakelam2015}. A recent comparison of five such codes against JWST ices in Chamaeleon~I \citep{jimenez2025} found that predicted abundances vary widely between models, and sensitivity tests show much of this depends on the BE values chosen \citep{Penteado_2017}. Our TPD test, in which the predicted CH$_3$OH BE distributions are propagated through the Polanyi--Wigner equation (Appendix~\ref{TPD}), shows the ML-predicted spectra reproduce the DFT peak desorption temperatures to within 10--12~K, supporting their direct use in distribution-aware models. Distribution-resolved BEs have been shown to shift snow-line locations and abundances relative to single-value treatments \citep{Bovolenta_2022,Grassi_2020, Bovolenta_2025, furuya2024, tinacci2023, bulik2026, He_2016}; our framework generates such distributions rapidly, at negligible cost compared with DFT sampling.
%%%%%%%%%%%%%%%%%%%%%%%%%%%%%%%%%%%%%%%%%%%%%%%%%%%%%%%%%%%%%%%%%%%%%%%%%%%%%
%%%%%%%%%%%%%%%%%%%%%%%%%%%%%%%%%%%%%%%%%%%%%%%%%%%%%%%%%%%%%%%%%%%%%%%%%%%%%
\section{Conclusions}
\label{conc}

This work demonstrates that BE distributions on amorphous water ice can be predicted directly from local geometric descriptors, achieving pooled LOCO performance of $R^2=0.90$ and MAE = 378 K without electronic-structure input. The model identifies a clear boundary between geometry-dominated adsorption, where H-bonding topology governs binding, and electronically driven regimes where dispersion forces or unpaired-electron interactions require information beyond local geometry.

Several directions follow from these results. Transfer to larger ice models would show whether the geometric correlations learned on ASW$_{12}$ surfaces generalise to the wider range of binding environments on extended amorphous ice. Expanding coverage to N- and S-bearing species is a priority, since NH$_3$, H$_2$S, and HNCO are prominent in JWST ice inventories \citep{McClure_2023}. Extending it to mixed ices (CO$_2$:H$_2$O, CO:H$_2$O, CH$_3$OH:H$_2$O) would reflect the multi-component nature of real mantles, where co-adsorbed species reshape the binding landscape. In the dispersion and radical regimes, where geometric descriptors reach their limits, low-cost electronic features (semi-empirical partial charges, HOMO/LUMO energies, spin-density indicators) offer a targeted way to improve accuracy without the full cost of DFT. Finally, coupling the model with active learning, which selects the most informative geometries for new DFT, would scale chemical and structural diversity efficiently.

These results show that local ice-surface geometry can provide sufficient information to predict BE distributions for H-bonding species, reducing the need for exhaustive quantum-chemical sampling in distribution-aware gas-grain models. As observational constraints on ice composition tighten, particularly through JWST programmes targeting interstellar and protoplanetary ices \citep{McClure_2023, smith2025, bergner2024}, rapid, physically grounded methods for generating BE distributions will become increasingly essential for interpreting the chemistry these observations reveal.
%%%%%%%%%%%%%%%%%%%%%%%%%%%%%%%%%%%%%%%%%%%%%%%%%%%%%%%%%%%%%%%%%%%%%%%%%%%%%
%%%%%%%%%%%%%%%%%%%%%%%%%%%%%%%%%%%%%%%%%%%%%%%%%%%%%%%%%%%%%%%%%%%%%%%%%%%%%
\begin{acknowledgments}
A.A. acknowledges support from UK Research and Innovation (grant numbers MR/T040726/1 and MR/Z00029X/1). C.W. acknowledges financial support from the Science and Technology Facilities Council and UK Research and Innovation (grant numbers ST/X001016/1, MR/T040726/1 and MR/Z00029X/1).
\end{acknowledgments}
%%%%%%%%%%%%%%%%%%%%%%%%%%%%%%%%%%%%%%%%%%%%%%%%%%%%%%%%%%%%%%%%%%%%%%%%%%%%%
%%%%%%%%%%%%%%%%%%%%%%%%%%%%%%%%%%%%%%%%%%%%%%%%%%%%%%%%%%%%%%%%%%%%%%%%%%%%%
\software{XGBoost \citep{chen2016}, scikit-learn \citep{pedregosa2011}, SHAP \citep{lundberg2017}, NumPy \citep{harris2020}, SciPy \citep{virtanen2020}, pandas \citep{mckinney2010}, Matplotlib \citep{hunter2007}}
%The complete training and evaluation pipeline, including the feature-extraction code, trained models, and scripts to reproduce all figures and tables, is publicly available at \url{https://github.com/<user>/<repo>}. 
The underlying BE dataset is described in \citet{ahmad2026}.
%%%%%%%%%%%%%%%%%%%%%%%%%%%%%%%%%%%%%%%%%%%%%%%%%%%%%%%%%%%%%%%%%%%%%%%%%%%%%
\bibliography{references}{}
\bibliographystyle{aasjournalv7}
%%%%%%%%%%%%%%%%%%%%%%%%%%%%%%%%%%%%%%%%%%%%%%%%%%%%%%%%%%%%%%%%%%%%%%%%%%%%%
%%%%%%%%%%%%%%%%%%%%%%%%%%%%%%%%%%%%%%%%%%%%%%%%%%%%%%%%%%%%%%%%%%%%%%%%%%%%%
\FloatBarrier
\appendix
\raggedbottom 
\section{Machine-learning model: supporting material}
\label{app:ml}

\begin{figure*}[tp]
    \centering
    \includegraphics[width=0.99\textwidth,height=0.85\textheight,keepaspectratio]{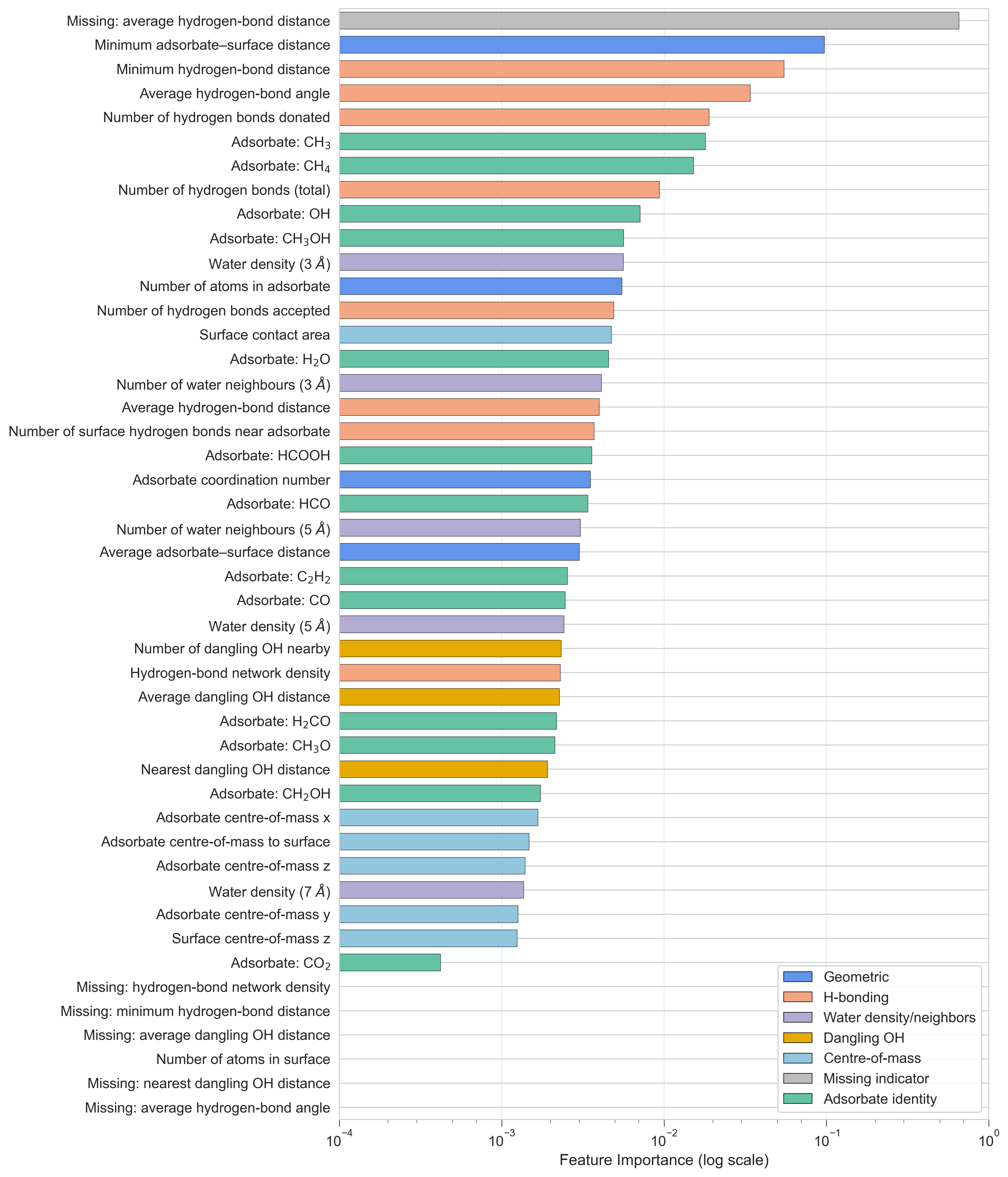}
    \caption{Gain-based feature importance for the 46 input features in the XGBoost model, ranked by decreasing importance. The dominant feature is the binary descriptor indicating that no hydrogen bond is formed, accounting for 63.4\% of the total gain. This descriptor is implemented as the indicator associated with an undefined average H-bond distance when no H-bond is present, and therefore has a direct physical interpretation rather than representing arbitrary missing data. The strongest continuous geometric predictors are the minimum adsorbate--surface distance (9.7\%) and the minimum H-bond distance (5.5\%). Five of the six binary H-bond indicators and the number of atoms in the surface have zero importance. Adsorbate identity features contribute only 7.1\% of the total gain, indicating that the model is driven primarily by properties of the local binding environment rather than species labels.}
    \label{fig:feature_importance_all}
\end{figure*}

\begin{figure*}[tp]
    \centering
    \includegraphics[width=0.99\textwidth,height=0.85\textheight,keepaspectratio]{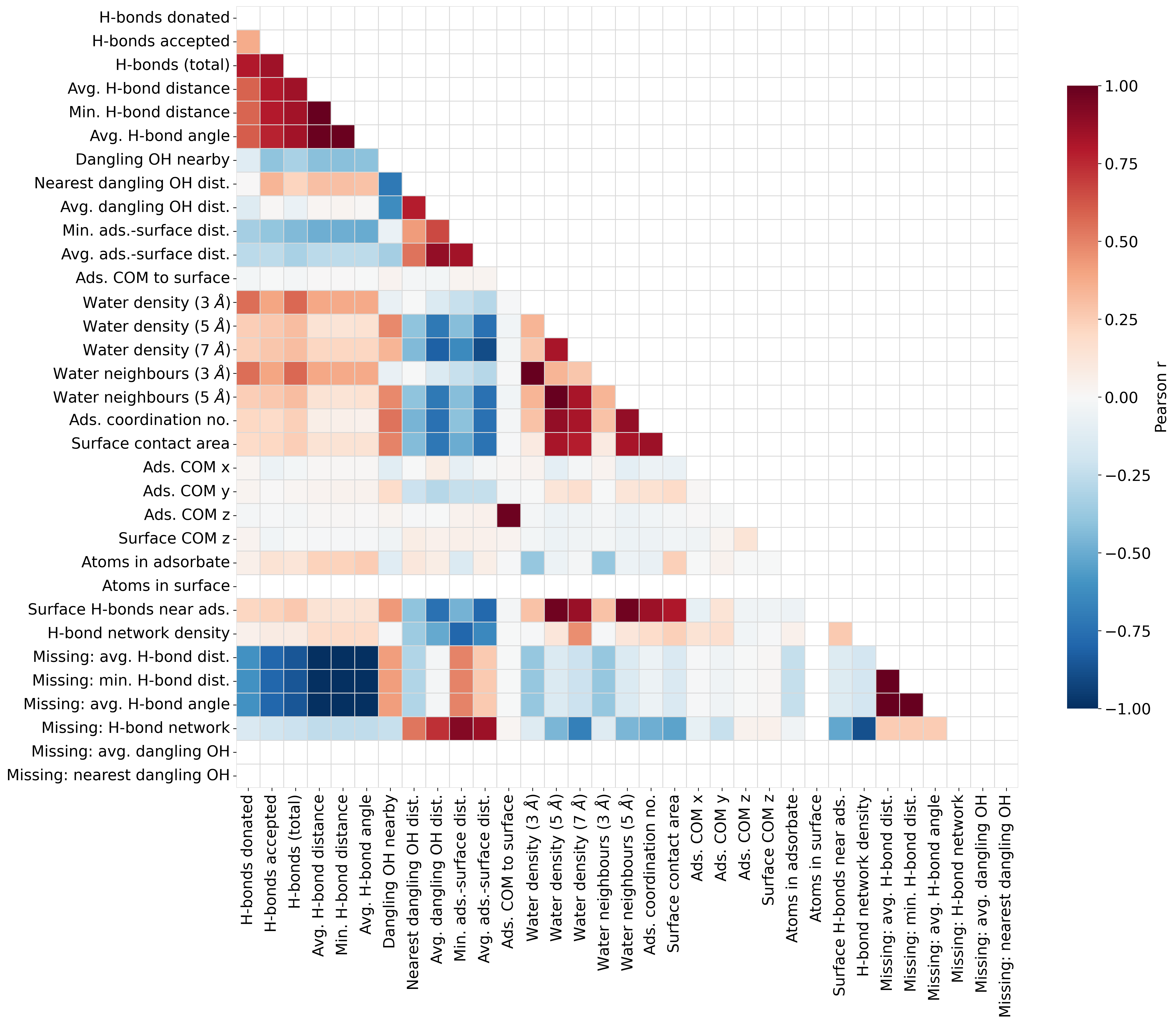}
    \caption{Pearson correlation matrix for the 33 non-categorical input features. The matrix covers the 27 geometric descriptors and 6 binary H-bond absence indicators used in the XGBoost model; the adsorbate one-hot columns are excluded, as they are uncorrelated by construction. Red marks positive correlation, blue negative, and white no linear relationship. Several blocks of strong correlation ($|r| > 0.8$) appear: (i) the three H-bond distance and angle descriptors correlate at $r > 0.99$, reflecting their shared dependence on H-bond formation; (ii) the six binary H-bond absence indicators correlate at $r = 1.00$, since each is triggered by the same physical condition, namely that no H-bond is formed; (iii) water density and water-neighbour count at matched radii are perfectly correlated ($r = 1.00$), since one is a linear rescaling of the other; and (iv) a broader block linking water density, coordination number, surface contact area, and nearby surface H-bonds ($r = 0.82$--$0.97$) reflects the local crowding of the binding site. In total, 47 feature pairs exceed $|r| = 0.8$. XGBoost is robust to correlated predictors for prediction, although such correlations can redistribute feature importance among redundant variables. In particular, the six H-bond absence indicators contain the same information, so the gain assigned to any one of them should be interpreted as representing the shared no-H-bond signal rather than that individual descriptor alone.}
    \label{fig:correl_matrix}
\end{figure*}

This appendix collects the supporting material for the machine learning model. Table~\ref{tab:feature_definitions} defines the 46 input features, Figure~\ref{fig:feature_importance_all} reports their gain-based importance, and Figure~\ref{fig:correl_matrix} shows the correlations among the non-categorical features. Table~\ref{tab:model_comparison} compares the regression methods benchmarked on the 80/20 cluster-based split, and Table~\ref{tab:hyperparam_sensitivity} summarises the sensitivity of XGBoost to its main hyperparameters. Table~\ref{tab:loco_clusters} gives the per-cluster leave-one-cluster-out results, Table~\ref{tab:learning_curve} the dependence of performance on training-set size, and Table~\ref{tab:uncertainty_calibration} the calibration of the bootstrap ensemble uncertainties.
%%%%%%%%%%%%%%%%%%%%%%%%%%%%%%%
%%%%%%%%%%%%%%%%%%%%%%%%%%%%%%%
\begin{table*}[tp]
\centering
\footnotesize
\caption{Definitions of the 46 input features used in the XGBoost model.}
\label{tab:feature_definitions}
\begin{tabular}{ll}
\hline
Feature & Definition \\
\hline
1. Minimum adsorbate--surface distance & Geometric; minimum adsorbate--surface atom distance; units: \AA \\
2. Average adsorbate--surface distance & Geometric; mean pairwise adsorbate--surface atom distance; units: \AA \\
3. Adsorbate coordination number & Geometric; mean number of surface atoms within 4~\AA \\
4. Number of atoms in adsorbate & Geometric; total atom count of adsorbate \\
5. Number of atoms in surface & Geometric; total atom count of water cluster \\
6. Minimum H-bond distance & H-bonding; shortest donor--acceptor H-bond distance; units: \AA \\
7. Average H-bond distance & H-bonding; mean donor--acceptor H-bond distance; units: \AA \\
8. Average H-bond angle & H-bonding; mean D--H$\cdots$A angle; units: degrees \\
9. Number of H-bonds donated & H-bonding; H-bonds donated by adsorbate \\
10. Number of H-bonds accepted & H-bonding; H-bonds accepted by adsorbate \\
11. Number of H-bonds total & H-bonding; sum of donated and accepted H-bonds \\
12. Surface H-bonds near adsorbate & H-bonding; nearby surface--surface H-bonds within 5~\AA \\
13. H-bond network density & H-bonding; nearby surface H-bonds divided by water neighbours within 5~\AA \\
14. Water density 3~\AA & Water density; local oxygen density within 3~\AA; units: atoms \AA$^{-3}$ \\
15. Water density 5~\AA & Water density; local oxygen density within 5~\AA; units: atoms \AA$^{-3}$ \\
16. Water density 7~\AA & Water density; local oxygen density within 7~\AA; units: atoms \AA$^{-3}$ \\
17. Water neighbours 3~\AA & Water density; surface oxygen atoms within 3~\AA \\
18. Water neighbours 5~\AA & Water density; surface oxygen atoms within 5~\AA \\
19. Dangling OH nearby & Dangling OH; dangling O--H groups within 5~\AA \\
20. Nearest dangling OH distance & Dangling OH; closest dangling OH hydrogen; units: \AA \\
21. Average dangling OH distance & Dangling OH; mean distance to dangling OH hydrogens; units: \AA \\
22. Surface contact area & Centre-of-mass; unique surface atoms within 3.5~\AA\ of adsorbate \\
23. Adsorbate COM x & Centre-of-mass; $x$-coordinate of adsorbate COM; units: \AA \\
24. Adsorbate COM y & Centre-of-mass; $y$-coordinate of adsorbate COM; units: \AA \\
25. Adsorbate COM z & Centre-of-mass; $z$-coordinate of adsorbate COM; units: \AA \\
26. Surface COM z & Centre-of-mass; $z$-coordinate of surface COM; units: \AA \\
27. Adsorbate COM to surface & Centre-of-mass; vertical adsorbate COM--surface distance; units: \AA \\
28. Missing average H-bond distance & Missing indicator; 1 if no H-bonds, 0 otherwise \\
29. Missing minimum H-bond distance & Missing indicator; 1 if no H-bonds, 0 otherwise \\
30. Missing average H-bond angle & Missing indicator; 1 if no H-bonds, 0 otherwise \\
31. Missing H-bond network density & Missing indicator; 1 if undefined, 0 otherwise \\
32. Missing average dangling OH distance & Missing indicator; 1 if no dangling OH groups, 0 otherwise \\
33. Missing nearest dangling OH distance & Missing indicator; 1 if no dangling OH groups, 0 otherwise \\
34. Adsorbate CH$_3$OH & Adsorbate identity; methanol one-hot feature \\
35. Adsorbate H$_2$O & Adsorbate identity; water one-hot feature \\
36. Adsorbate CO & Adsorbate identity; carbon monoxide one-hot feature \\
37. Adsorbate HCO & Adsorbate identity; formyl radical one-hot feature \\
38. Adsorbate OH & Adsorbate identity; hydroxyl radical one-hot feature \\
39. Adsorbate CO$_2$ & Adsorbate identity; carbon dioxide one-hot feature \\
40. Adsorbate H$_2$CO & Adsorbate identity; formaldehyde one-hot feature \\
41. Adsorbate CH$_3$O & Adsorbate identity; methoxy radical one-hot feature \\
42. Adsorbate CH$_2$OH & Adsorbate identity; hydroxymethyl radical one-hot feature \\
43. Adsorbate HCOOH & Adsorbate identity; formic acid one-hot feature \\
44. Adsorbate CH$_4$ & Adsorbate identity; methane one-hot feature \\
45. Adsorbate C$_2$H$_2$ & Adsorbate identity; acetylene one-hot feature \\
46. Adsorbate CH$_3$ & Adsorbate identity; methyl radical one-hot feature \\
\hline
\end{tabular}

\vspace{0.5em}
\begin{minipage}{0.96\textwidth}
\footnotesize
\textbf{Notes.} H-bond detection criteria: donor--acceptor distance $\leq 3.5$~\AA\ and D--H$\cdots$A angle $\geq 120^\circ$. Dangling OH is defined as a surface O--H group where the hydrogen is not within 2.5~\AA\ of another oxygen. Centre of mass is computed using atomic masses. Missing values are imputed with 0.0 prior to model training.
\end{minipage}

\end{table*}
%%%%%%%%%%%%%%%%%%%%%%%%%%%%%%%%%%%%%%%%%
\begin{table}[htbp]
\centering
\small
\setlength{\tabcolsep}{8pt}
\renewcommand{\arraystretch}{1.15}

\caption{Sensitivity of XGBoost performance to key hyperparameters on the 80/20 cluster-based train/test split. Test-set performance is reported as $R^2$ and MAE in K.}
\label{tab:hyperparam_sensitivity}

\begin{tabular}{@{}llcc@{}}
\toprule
\textbf{Varied parameter} & \textbf{Value} & \textbf{Test $R^2$} & \textbf{MAE (K)} \\
\midrule
\multirow{4}{*}{\texttt{max\_depth}}
  & 3  & 0.886 & 392 \\
  & 5$^\ast$ & 0.891 & 387 \\
  & 7  & 0.893 & 385 \\
  & 9  & 0.892 & 386 \\
\addlinespace[3pt]

\multirow{4}{*}{\texttt{n\_estimators}}
  & 50   & 0.884 & 394 \\
  & 100$^\ast$ & 0.891 & 387 \\
  & 200  & 0.893 & 384 \\
  & 500  & 0.894 & 383 \\
\addlinespace[3pt]

\multirow{4}{*}{\texttt{learning\_rate}}
  & 0.01 & 0.870 & 406 \\
  & 0.05 & 0.889 & 389 \\
  & 0.1$^\ast$ & 0.891 & 387 \\
  & 0.3  & 0.886 & 393 \\
\bottomrule
\end{tabular}

\vspace{0.3em}
\parbox{0.95\columnwidth}{\footnotesize\textbf{Notes.} \texttt{max\_depth} is the
maximum number of splits in a single tree, setting how complex a pattern it can
fit; \texttt{n\_estimators} is the number of trees in the ensemble; and
\texttt{learning\_rate} sets how strongly each new tree adjusts the running
prediction, smaller values training more gradually. Asterisks denote the baseline
values in the reference XGBoost configuration. All results use the same 80/20
cluster-based train/test split.}

\end{table}
%%%%%%%%%%%%%%%%%%%%%%%%%%%%%%%%%%%%%%%%%
\begin{table}[htbp]
\centering
\small
\setlength{\tabcolsep}{7pt}
\renewcommand{\arraystretch}{1.15}

\caption{Performance comparison of machine-learning methods on the 80/20 cluster-based train/test split. $R^2$ is the coefficient of determination, the fraction of variance in the binding energies explained by the model, where $R^2 = 1$ is a perfect fit and $R^2 = 0$ is no better than predicting the mean. RMSE and MAE are reported in K.}
\label{tab:model_comparison}

\begin{tabular}{@{}lccccc@{}}
\toprule
\textbf{Method} & \textbf{Train $R^2$} & \textbf{Test $R^2$} & \textbf{RMSE (K)} & \textbf{MAE (K)} & \textbf{Overfit gap} \\
\midrule
Linear regression   & 0.898 & 0.887 & 606 & 424 & 0.011 \\
Ridge regression    & 0.892 & 0.882 & 619 & 425 & 0.010 \\
Random forest       & 0.979 & 0.885 & 609 & 406 & 0.094 \\
Gradient boosting   & 0.990 & 0.886 & 607 & 387 & 0.104 \\
XGBoost (baseline)  & 0.987 & 0.891 & 593 & 387 & 0.095 \\
XGBoost (tuned)     & 0.990 & 0.894 & 585 & 383 & 0.096 \\
\midrule
\multicolumn{6}{@{}l@{}}{\textit{Additional kernel-method benchmark (test metrics only)}} \\
XGBoost (baseline, benchmark) & -- & 0.901 & -- & 380 & -- \\
GPR (RBF + White)             & -- & 0.874 & -- & 454 & -- \\
Kernel ridge (RBF)            & -- & 0.840 & -- & 457 & -- \\
\midrule
LOCO-CV (pooled) & -- & 0.896 & 574 & 378 & -- \\
\bottomrule
\end{tabular}

\vspace{0.3em}
\parbox{\textwidth}{\footnotesize\textbf{Notes.} The overfit gap is the training $R^2$ minus the test $R^2$. RMSE is the root-mean-square error: the typical size of a prediction error, in K, with larger errors weighted more heavily than small ones. The tuned XGBoost hyperparameters were selected by randomised-search cross-validation (80 iterations, five-fold). Gaussian process regression (GPR) and kernel ridge regression (KRR) both use a radial basis function kernel fitted to standardised features. LOCO-CV (pooled) reports metrics from leave-one-cluster-out cross-validation across all 15 ASW$_{12}$ clusters. All single-split methods use the same cluster-based partition (seed $= 42$; 12 training clusters and 3 test clusters). XGBoost outperforms the kernel methods by 3--6 percentage points in $R^2$, likely reflecting the inability of a single global length scale to capture the heterogeneous binding regimes present in the data. For reference, 1~kJ\,mol$^{-1}$ = 120.27~K.}

\end{table}
%%%%%%%%%%%%%%%%%%%%%%%%%%%%%
%%%%%%%%%%%%%%%%%%%%%%%%%%%%%%
\begin{table}[htbp]
\centering
\small
\setlength{\tabcolsep}{5pt}
\renewcommand{\arraystretch}{1.15}

\caption{Leave-one-cluster-out (LOCO) cross-validation results for each ASW$_{12}$ cluster. $N$ is the number of binding structures in the held-out cluster. $R^2$ is the coefficient of determination; RMSE and MAE are reported in K. Clusters are ordered by fold index (the order in which each cluster was held out as the validation fold), which is arbitrary with respect to $N$, $R^2$, RMSE, and MAE.}
\label{tab:loco_clusters}

\begin{tabular}{@{}lcccc@{}}
\toprule
\textbf{Cluster} & \textbf{$N$} & \textbf{$R^2$} & \textbf{RMSE (K)} & \textbf{MAE (K)} \\
\midrule
ASW$_{12}$\_01  & 23  & 0.944  & 461  & 348 \\
ASW$_{12}$\_02  & 83  & 0.949  & 415  & 291 \\
ASW$_{12}$\_03  & 77  & 0.973  & 321  & 199 \\
ASW$_{12}$\_04  & 147 & 0.793  & 947  & 677 \\
ASW$_{12}$\_05  & 86  & 0.933  & 440  & 359 \\
ASW$_{12}$\_06  & 212 & 0.922  & 454  & 291 \\
ASW$_{12}$\_10 & 123 & 0.852  & 723  & 449 \\
ASW$_{12}$\_11 & 120 & 0.912  & 499  & 322 \\
ASW$_{12}$\_13 & 11  & $-2.799$ & 765  & 683 \\
ASW$_{12}$\_14 & 121 & 0.928  & 452  & 330 \\
ASW$_{12}$\_15 & 126 & 0.901  & 528  & 376 \\
ASW$_{12}$\_16 & 126 & 0.930  & 425  & 301 \\
ASW$_{12}$\_18 & 90  & 0.911  & 430  & 294 \\
ASW$_{12}$\_19 & 67  & 0.844  & 594  & 429 \\
ASW$_{12}$\_20 & 17  & -0.097 & 1284 & 1145 \\
\midrule
\textbf{Pooled} & \textbf{1429} & \textbf{0.90} & \textbf{574} & \textbf{378} \\
\bottomrule
\end{tabular}

\vspace{0.3em}
\parbox{0.95\columnwidth}{\footnotesize\textbf{Notes.} Pooled metrics are computed from all 1,429 predictions aggregated across the 15 folds. The negative $R^2$ values for ASW$_{12}$\_13 ($N = 11$) and ASW$_{12}$\_20 ($N = 17$) reflect the small sample sizes, which provide insufficient local variation for meaningful regression evaluation. Cluster labels follow the numbering used in \citealt{ahmad2026}.}

\end{table}
%%%%%%%%%%%%%%%%%%%%%%%%%%%%%%%
\begin{table}[htbp]
\centering
\small
\setlength{\tabcolsep}{6pt}
\renewcommand{\arraystretch}{1.15}

\caption{Model performance as a function of training-set size. Leave-one-cluster-out (LOCO) cross-validation $R^2$ and MAE are reported for subsets of the ASW$_{12}$ cluster pool.}
\label{tab:learning_curve}

\begin{tabular*}{\linewidth}{@{\extracolsep{\fill}} l l l l @{}}
\toprule
\textbf{Clusters} & \textbf{Structures} & \textbf{$R^{2}$ (mean $\pm$ std)} & \textbf{MAE (K)} \\
\midrule
5  & $\sim$490  & $0.878 \pm 0.039$ & $421 \pm 82$ \\
7  & $\sim$720  & $0.899 \pm 0.023$ & $371 \pm 50$ \\
9  & $\sim$960  & $0.891 \pm 0.011$ & $385 \pm 27$ \\
11 & $\sim$1090 & $0.894 \pm 0.014$ & $382 \pm 36$ \\
13 & $\sim$1290 & $0.892 \pm 0.001$ & $388 \pm 3$ \\
15 & 1429       & $0.896$           & $378$ \\
\bottomrule
\end{tabular*}

\vspace{0.3em}
\parbox{0.95\columnwidth}{\footnotesize\textbf{Notes.} LOCO cross-validation results are computed on subsets of the ASW$_{12}$ cluster pool, with the number of training clusters increasing from 5 to 15. For each subset size, five random draws of clusters were evaluated and the mean and standard deviation are reported. The final row (15 clusters) uses the full dataset with no resampling. $R^2$ plateaus at 0.89--0.90 from seven clusters onward, indicating that the dataset of 1,429 structures across 15 clusters is sufficient and that additional clusters would likely yield diminishing returns.}

\end{table}
%%%%%%%%%%%%%%%%%%%%%%%%%%%%%%%
\FloatBarrier

\begin{table}[htbp]
\centering
\small
\setlength{\tabcolsep}{10pt}
\renewcommand{\arraystretch}{1.15}

\caption{Bootstrap ensemble uncertainty calibration. Test-set structures are grouped into quintiles by predicted uncertainty, $\sigma$, defined as the standard deviation across 50 XGBoost models trained on cluster-level bootstrap resamples of the training set.}
\label{tab:uncertainty_calibration}

\begin{tabular}{@{}lcc@{}}
\toprule
\textbf{Bin} & \textbf{Mean $\sigma$ (K)} & \textbf{Observed RMSE (K)} \\
\midrule
1 & 96  & 268 \\
2 & 151 & 517 \\
3 & 209 & 583 \\
4 & 277 & 611 \\
5 & 390 & 706 \\
\bottomrule
\end{tabular}

\vspace{0.3em}
\parbox{0.95\columnwidth}{\footnotesize\textbf{Notes.} The test set contains $n = 267$ structures. The observed RMSE increases monotonically with predicted $\sigma$ (Spearman $\rho = 0.365$, $p = 1.4 \times 10^{-12}$), confirming that the ensemble uncertainty estimates are informative: structures assigned higher uncertainty by the model genuinely exhibit larger prediction errors. Unlike Gaussian-process regression posterior variance, which assumes a stationary covariance structure, bootstrap ensemble variance is distribution-free and makes no assumptions about the functional form of the noise.}
\end{table}
%%%%%%%%%%%%%%%%%%%%%%%%%%%%%%%%%%
\clearpage
\section{TPD Simulation}
\label{TPD}

TPD spectra were simulated from the BE
distributions using the Polanyi--Wigner equation \citep{polanyi1928} for
first-order desorption:
\begin{equation}
  \frac{\mathrm{d}N}{\mathrm{d}T}
  = -\frac{\nu}{\beta}\, N \exp\!\left(-\frac{E_\mathrm{des}}{k_\mathrm{B} T}\right),
\end{equation}
where $N$ is the surface population (the number of molecules still adsorbed), $T$ is the temperature, and $\mathrm{d}N/\mathrm{d}T$ is the resulting desorption rate, which constitutes the simulated TPD signal. The pre-exponential factor is $\nu$, $\beta$ is the heating rate (1~K\,min$^{-1}$), $E_\mathrm{des}$ is the desorption energy per molecule, and $k_\mathrm{B}$ is the Boltzmann constant. Each binding site was treated as an independent first-order desorption channel, and the total desorption rate was summed over all sites at each temperature step.

Since the DFT and ML datasets provide a discrete set of approximately 15--25 BEs per adsorbate per surface, kernel density estimation (a Gaussian KDE with
Scott's rule for bandwidth selection; \citealt{scott2011}) was used to construct a continuous BE distribution from the discrete data points. Five hundred virtual binding sites were resampled from this distribution, and resampled values were clipped to within 20\% of the observed data range to exclude unphysical KDE tail artefacts. Structures with BEs below 2~kJ\,mol$^{-1}$ ($\approx 240$~K) were excluded as essentially unbound. The same procedure and parameters were applied identically to both DFT and ML-predicted BEs.

\begin{figure*}[t]
  \centering
  \includegraphics[width=\textwidth]{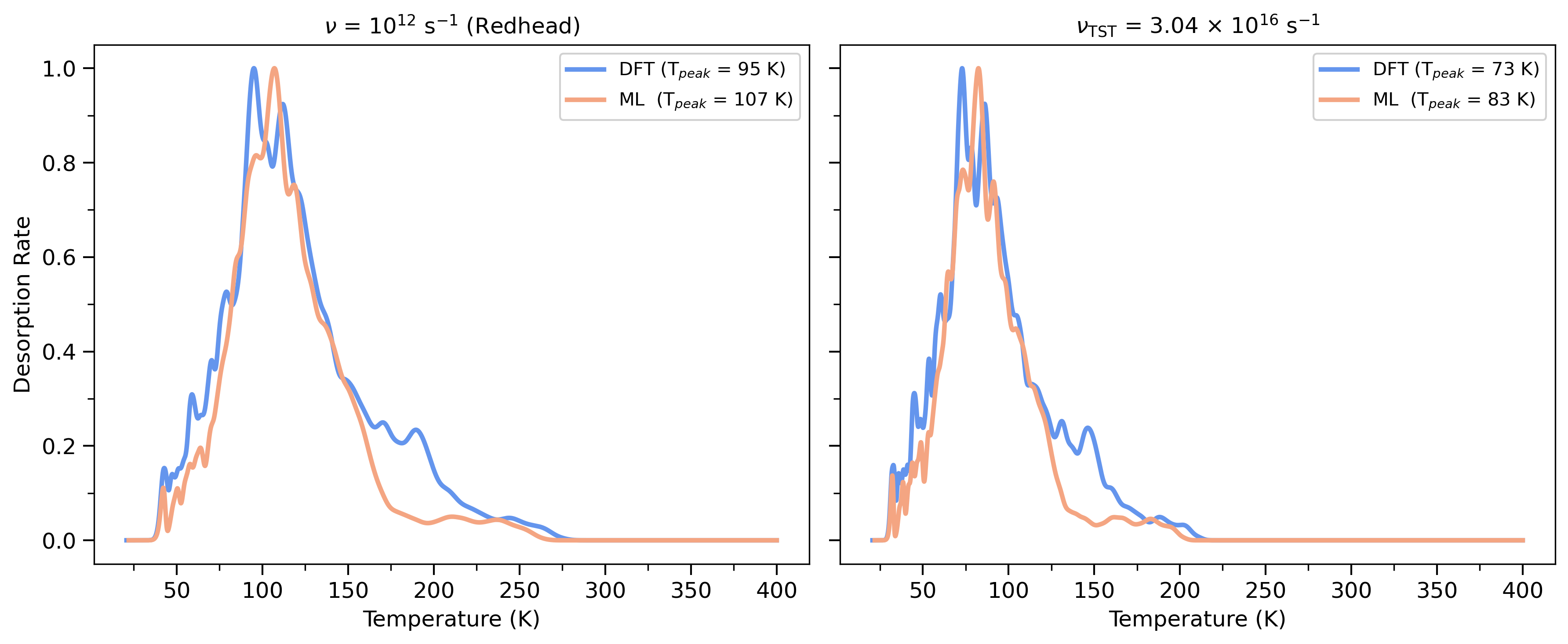}
  \caption{Simulated TPD spectra for CH$_3$OH on amorphous water ice. Desorption rate curves computed from DFT binding energies and ML-predicted binding energies using the Polanyi--Wigner equation with 500 KDE-resampled virtual sites. Left: Redhead pre-exponential factor ($\nu = 10^{12}\ \mathrm{s^{-1}}$). Right: species-specific transition-state-theory pre-exponential factor ($\nu_{\mathrm{TST}} = 3.04 \times 10^{16}\ \mathrm{s^{-1}}$, \citealt{ahmad2026}).}
  \label{fig:TPD}
\end{figure*}

Two pre-exponential factors were employed for the detailed CH$_3$OH case study: the standard Redhead approximation \citep{redhead1962} ($\nu = 10^{12}$~s$^{-1}$), and a transition-state-theory value computed at 10~K by \citet{ahmad2026} ($\nu_\mathrm{TST} = 3.04 \times 10^{16}$~s$^{-1}$).
%%%%%%%%%%%%%%%%%%
To demonstrate that the predicted BEs yield physically meaningful observables, we simulated TPD spectra for CH$_3$OH using the Polanyi--Wigner equation, with BEs drawn from the DFT and ML-predicted distributions (Appendix Figure~\ref{fig:TPD}). Using a Redhead pre-exponential factor ($\nu = 10^{12}$~s$^{-1}$) and a heating rate of $\beta = 0.017$~K\,s$^{-1}$ (1~K\,min$^{-1}$), the DFT and ML spectra yield peak desorption temperatures of 95~K and 107~K, respectively. With a species-specific transition-state-theory pre-exponential factor for CH$_3$OH ($\nu_\mathrm{TST} = 3.04 \times 10^{16}$~s$^{-1}$; \citealt{ahmad2026}), the corresponding peaks shift to 73~K and 83~K. In both cases the ML-predicted spectral shape closely follows the DFT reference, with peak-temperature differences of 10--12~K, confirming that the model produces BE distributions consistent with observable desorption behaviour.
%%%%%%%%%%%%%%%%%%%
%%%%%%%%%%%%%%%%%%%%%%%%%%%%%%%%%%%%%%%%%%%%%%%%%%%%%%%%%%%%%%%%%%%%%%%%%%%%%
\end{document}